# An Atypical Survey of Typical-Case Heuristic Algorithms[*]

Lane A. Hemaspaandra
Department of Computer Science
University of Rochester
Rochester, NY 14627, USA[†]

Ryan Williams
Computer Science Department
Stanford University
Stanford, CA 94305, USA[‡]

October 30, 2012

**Abstract**

Heuristic approaches often do so well that they seem to pretty much always give the right answer. How close can heuristic algorithms get to always giving the right answer, without inducing seismic complexity-theoretic consequences? This article first discusses how a series of results by Berman, Buhrman, Hartmanis, Homer, Longpré, Ogiwara, Schöning, and Watanabe, from the early 1970s through the early 1990s, explicitly or implicitly limited how well heuristic algorithms can do on NP-hard problems. In particular, many desirable levels of heuristic success cannot be obtained unless severe, highly unlikely complexity class collapses occur. Second, we survey work initiated by Goldreich and Wigderson, who showed how under plausible assumptions deterministic heuristics for randomized computation can achieve a very high frequency of correctness. Finally, we consider formal ways in which theory can help explain the effectiveness of heuristics that solve NP-hard problems in practice.

## 1 Introduction

Your boss sends you an email inviting you to stop by his office. Great. As you pay him a visit, your boss says, "No pressure at all, but I'd really like it if you could help me by providing a polynomial-time (deterministic) algorithm for this problem." You notice that your boss is holding a mostly blank document that is your yearly performance review and salary recommendation. Uh-oh. Maybe this isn't just a friendly chat? No problem—all you have to do is find a polynomial-time algorithm. Then you notice that the problem is NP-hard. After a moment of terror you—like many researchers before you—recall the famed page 3 of the classic book of Garey and Johnson [GJ79], and you explain to your boss that due to the NP-hardness, "I can't find an efficient algorithm, but neither can all these famous people."

[*]This article also appears as University of Rochester Department of Computer Science Technical Report TR-2012-984, and is scheduled to appear in the December 2012 issue of *SIGACT News*.

[†]`www.cs.rochester.edu/u/lane`. Supported in part by NSF grants CCF-0915792 and CCF-1101479.

[‡]`www.stanford.edu/~rrwill`. Supported in part by a David Morgenthaler II Faculty Fellowship and NSF grant CCF-1212372.

Breathing a sigh of relief, you start to get up, but you notice that your boss is glancing down again at the performance-review form. Gulp. Says your boss, "Ah—excellent point. But I'd really like it if you could at least provide me with a polynomial-time algorithm that is almost never wrong." You ask him just how "almost never wrong" he expects your algorithm for the problem to be, and he thinks and replies, "hmm, let's say, it should be wrong on at most polynomially many strings at each length $n$."

Uh-oh, again. You know every line of Garey and Johnson and frantically start reviewing those in your mind, but that's not helping.

Now, don't be too hard on your boss. The tremendous success of heuristic algorithms has led many people to believe that heuristic algorithms can leap tall buildings in a single bound. At a recent Ph.D. thesis defense one of us attended, where the thesis was in part about heuristics for NP-hard problems, the candidate was asked what bound he thought held for the error frequency of one of his heuristic algorithms, and was given some options, and although he was cautious enough not to claim polynomial error frequency, he said that he believed that his algorithm would err on no more than "$n$ to the polylog($n$)" strings per length—a less demanding degree of success than your boss is expecting. The student didn't realize that his belief, if true, would change the face of complexity theory by giving unexpectedly fast deterministic algorithms for all of NP—so fast that anyone who could establish that would likely have a Turing Award coming in the mail.

But your boss is still contemplating that performance review. You think, fast and hard, and then (through remembering the relevant result, or through a burst of insight or on-the-spot theorem-proving) give the following correct reply. You explain to your boss that due to the NP-hardness, "I can't find the excellent heuristic algorithm you seek, but neither can all these famous people." Your boss's second request is in fact just as unlikely to be doable as is your boss's first request; satisfying the second request would cause NP to equal P.

This article is organized as follows. Section 2 covers a number of complexity results, from the early 1970s through the early 1990s, that limit or can be used to limit how often heuristic algorithms can be correct on NP-hard problems. These results were obtained by Berman, Buhrman, Hartmanis, Homer, Longpré, Ogiwara, Schöning, and Watanabe. (In 2008, Buhrman and Hitchcock proved results that extend part of this line, and that we also discuss.) Particularly remarkable in this bit of history is how complexity-theoretic work from the 20th century sharply limits the performance of heuristic algorithms in the 21st century. Heuristic algorithms potentially can do quite well—but they can't ever do well on almost all instances of an NP-hard problem unless all of NP has surprisingly fast algorithms (or the polynomial hierarchy collapses). Section 3 discusses a similar challenge in the realm of P versus BPP: namely, finding deterministic heuristic algorithms, for problems solvable with randomness, that are correct on all but a small number of inputs. In this setting, the known results are generally positive. Often, such heuristic algorithms can actually be found, but we do not know when or whether these heuristics imply fully correct algorithms. Nevertheless, these heuristics add to the evidence that P = BPP. Finally, Section 4 discusses the surprising effectiveness of heuristic algorithms for solving practical instances of NP-hard problems, and considers theoretical models for explaining what makes practical instances different from the "worst-case" instances discussed in the earlier sections.

Our goal in this article is not to prove the results involved, but rather to highlight them (thus perhaps helping future people with insistent bosses), to give pointers to the original sources, to explain why the results do in fact limit how well heuristics can do, and to pose scientific questions about real-world instances that can be both studied theoretically and verified (or falsified) experimentally. We stress that this article is not at all an attempt to cover everything that is known about heuristic algorithms, or even to cover a decent fraction of the more important results. We are mainly focused on so-called "typical-case hardness" rather than Levin-style average-case hardness, and we even within typical-case hardness focus on frequency counts and models of real-world problems, rather than on probabilistic distributions over the inputs. We highly recommend to readers wishing a broader, deeper coverage of issues related to heuristic algorithms and typical-case complexity the monograph of Bogdanov and Trevisan [BT06] and the references therein.

## 2 Quite Old Complexity Results that Limit the Performance of Heuristic Algorithms

Throughout this article, when we speak of a heuristic algorithm, we will be speaking about deterministic algorithms. (That is, we will not be speaking of the case where the heuristic algorithm is *randomized*, and is trying to achieve a low asymptotic *expected* error frequency.) So for us a polynomial-time heuristic algorithm will simply be a (deterministic) polynomial-time program, and of course it will accept some P set. (We will also briefly discuss later so-called "errorless" heuristic algorithms, which can answer *Yes*, *No*, or *Don't Know*.)

In this section and the next, we will focus on what limits hold for the asymptotic error frequency. That is, suppose we are interested in finding good heuristic algorithms for a set $L$. Suppose we are considering some polynomial-time heuristic algorithm, $\mathcal{A}$. As mentioned above, the algorithm itself is accepting some set of strings, in particular, some P set, call it $L_{\mathcal{A}}$. The error frequency, $\text{err}(n)$, is the number of instances at length $n$ on which the heuristic algorithm errs, i.e., $\text{err}(n) = (L \cap \Sigma^n) \triangle (L_{\mathcal{A}} \cap \Sigma^n)$, where $\triangle$ denotes the symmetric difference, $X \triangle Y = (X - Y) \cup (Y - X)$. (One can also define an error frequency based on the number of errors that occur at all lengths up to and including $n$. For the density levels we consider, the difference is usually not an interesting difference. For example, a set has polynomial (respectively, $n^{\log^{\mathcal{O}(1)}(n)}$, i.e., "$n$ to the polylog$(n)$") error frequency under one definition if and only if it has polynomial (respectively, $n^{\log^{\mathcal{O}(1)}(n)}$) error frequency under the other definition.)

## 2.1 Low Error Frequency and Speeding-Up-SAT Consequences

### 2.1.1 Polynomial Error Frequency

Probably the most important result regarding whether heuristic algorithms for SAT can have low error frequency is a theorem sought by Schöning in the 1980s [Sch86] and then achieved by Ogihara and Watanabe [OW91].

First, a few definitions of standard complexity notions will be helpful. A sparse set is a set such that there exists a polynomial $p$ such that for each $n$ the set contains at most $p(n)$ elements of length $n$. A set $A$ is said to be P-close if there exists some set $B \in$ P such that $A \triangle B$ is sparse ([Sch86], see also [Yes83]). A set $A$ polynomial-time $k$-truth-table reduces to $B$ if $A$ reduces to $B$ by a polynomial-time Turing reduction that on each input makes at most $k$ queries to its oracle $B$ and makes them all in parallel [LLS75]. A set $A$ polynomial-time bounded-truth-table reduces to $B$ if there is an integer $k$ such that $A$ polynomial-time $k$-truth-table reduces to $B$ [LLS75].

Schöning proved that if any paddable NP-hard (by which in this article we will always mean NP-$\leq_m^p$-hard unless we explicitly state a different type of hardness) set is P-close, then P = NP. We do not have to define here what it means to be paddable, since that restriction was in effect removed by a later result. In particular, Ogiwara and Watanabe [OW91] proved that if any sparse set is polynomial-time bounded-truth-table-hard for NP, then P = NP. This important result allows one to remove the "paddable" requirement from Schöning's theorem, since clearly every P-close set polynomial-time *one*-truth-table reduces to a sparse set.[1]

So the following result holds as an immediate consequence of Ogiwara and Watanabe's result.

**Theorem 2.1 (Follows from [OW91])** *If any NP-hard set is P-close, then* P = NP.

Of course, the above result applies directly to heuristic algorithms. In fact, Schöning [Sch86] and the related earlier work of Yesha [Yes83] are both inspired in part by this connection to heuristic algorithms. The above result, in particular, is just another way of saying the following.

**Corollary 2.2 (Follows from [OW91])** *No polynomial-time heuristic algorithm for any NP-hard problem can have a polynomial error frequency (i.e., can have* err(n) *be polynomially bounded), unless* P = NP.

Note how powerfully this work from more than twenty years ago addresses whether any polynomial-time heuristic algorithm can have a low error frequency on any NP-complete set: If even one polynomial-time heuristic algorithm has a polynomial error frequency on even one NP-complete problem, then all NP-complete sets are in P anyway, so one doesn't really need heuristic algorithms for them, since one can just solve them, exactly, in deterministic polynomial time.

---

[1]We mention in passing that each P-close set is, obviously, in the class P/poly, i.e., it has small circuits. And so if any NP-hard set is P-close, the polynomial hierarchy collapses to NP$^{\text{NP}}$ [KL80] and indeed to the "symmetric alternating" class $S_2^p$ (see [Cai01]). But this is of course a much less dramatic collapse than something like P = NP.

### 2.1.2 Subexponential Error Frequency

As described above, by the very early 1990s there was a stake in the heart of the possibility that any heuristic algorithm for any NP-hard problem could have an extremely low error frequency. But hope springs eternal. The natural next step is to allow vastly more errors. In particular, a natural and much more forgiving amount of error to allow would be to allow quasi-polynomially many errors—that is, to ask that $\text{err}(n) = n^{\log^{\mathcal{O}(1)}(n)}$. This would allow error frequencies such as $n^{\log n}$ and even $2^{\log^{2012}(n)}$.

That hope had a very short lifespan. Homer and Longpré [HL94] looked hard at the proof of Ogiwara and Watanabe [OW91], and how to improve the consequences and constants involved in certain cases, and Buhrman and Homer [BH92], building on that (the inverted dates are a conference-versus-journal effect), noted that it is highly unlikely that there exists any set that is bounded-truth-table-hard for NP (or bounded-truth-table complete) and whose census function (number of strings per length, or number of strings up to a given length; for our purposes both definitions work equally well, as noted earlier) is quasi-polynomially bounded. In particular, if there exists a set that is bounded-truth-table-hard for NP and whose census function is quasi-polynomially bounded, then all NP sets are in quasi-polynomial time ($\text{DTIME}[n^{\log^{\mathcal{O}(1)}(n)}]$), and nondeterministic exponential time equals deterministic exponential time (i.e., $\text{DTIME}[2^{n^{\mathcal{O}(1)}}] = \text{NTIME}[2^{n^{\mathcal{O}(1)}}]$ or, equivalently, EXP = NEXP) and if one further assumes that the set is not merely NP bounded-truth-table-hard but in fact is even NP bounded-truth-table-complete, then one additionally can conclude that the so-called (weak) exponential hierarchy of Hartmanis, Immerman, and Sewelson [HIS85] collapses [HL94, BH92].

However, and our comments on this have been earlier pointed out by Faliszewski, Hemaspaandra, and Hemaspaandra [FHH11], that work, analogously to the sparseness case, puts a huge roadblock in the way of any hope that polynomial-time heuristic algorithms having quasi-polynomial error frequency, can exist for any NP-hard problem. The reason is simply that if you have a polynomial-time heuristic algorithm for a problem, and that algorithm has only quasi-polynomial error frequency, then clearly the problem polynomial-time *one*-truth-table reduces to a set having a quasi-polynomially bounded census function (for example, the most natural such set is the set of instances on which the heuristic algorithm errs). So, in particular:

> *If some polynomial-time heuristic algorithm for an NP-hard set has quasi-polynomially bounded error frequency, then all NP sets are in quasi-polynomial time and nondeterministic exponential time equals deterministic exponential time.*

Put another way, if anyone tells you that he or she has a polynomial-time heuristic algorithm for some NP-hard problem and has proven its error frequency to be quasi-polynomially bounded, shake that person's hand, because that person (unless mistaken) will soon have a Turing Award.

The natural next question to ask is: do these same strong consequences still hold assuming heuristics with even higher error frequencies? Could we have deterministic polynomial-time heuristics for NP problems that make at most $2^{n^{o(1)}}$ errors? Alas, this also appears unlikely. In 2008, Buhrman and Hitchcock [BH08] proved the following.

**Theorem 2.3** *If there is a set $S$ such that*

*(a) for all $\varepsilon > 0$, the number of strings in $S$ of length $n$ is asymptotically at most $2^{n^{\varepsilon}}$, and*

*(b) there is an* NP*-hard set that polynomial-time bounded-truth-table reduces to $S$,*

*then* NP $\subseteq$ coNP/poly.

That is, if there are NP-hard sets that are "subexponentially dense," then NP $\subseteq$ coNP/poly and so the polynomial hierarchy collapses (all the way down to the class $S_2^{NP}$, by a result of Cai, Chakaravarthy, Hemaspaandra, and Ogihara [CCHO05]). By the above discussion, this implies that no polynomial-time heuristic for an NP-hard problem can make only subexponentially errors, unless another amazing complexity-theoretic consequence happens. This error bound is essentially optimal, as we'll see in the next section.

Finally, we note that there is a different approach to heuristic algorithms, often called "errorless" heuristic algorithms (see, e.g., [BS07] and the references therein), in which the algorithm must always be correct when it says *Yes* or *No*, but can also choose to sometimes say *Don't Know*. We mention in passing that that approach is well-covered by the results mentioned in this section and Section 2.1.1. As long as that algorithm runs in polynomial time, and as long as the frequency of *Don't Know*s is at most subexponential (to get one or more of the consequences of this section) or at most polynomial (to get the stronger consequences given in Section 2.1.1), then consequences discussed in this section or the previous one apply. Indeed this is an easier case; the results already discussed are stronger, in that they apply even to heuristics whose errors' locations are not pointed out by the heuristic algorithm.

## 2.2 Padding, and What Error Frequencies Can We Expect?

Section 2.1 discusses work showing that the existence of a polynomial-time heuristic algorithm for even one NP-hard problem would—if the algorithm had a polynomially bounded or quasi-polynomially bounded error frequency—collapse nondeterministic and deterministic computation classes, yielding P = NP or EXP = NEXP. Proving either of those equalities would be an amazing achievement, if they are even true.

What error frequencies can one hope to obtain without having to make Turing-Award-level breakthroughs in the process? By using a well-known "padding" trick, for each $k$ it is easy to construct NP-complete problems that have polynomial-time heuristic algorithms whose error frequency is $2^{n^{1/k}}$. In particular, fixing an arbitrary $k$, each NP-hard set $A$ is polynomial-time many-one reducible to—indeed, is even polynomial-time many-one equivalent to—the set $A' = \{x1^{|x|^k + k}\}$ (see the last part of Section 2 of [EHRS09]), and these sets clearly have the desired error-frequency bounds with respect to the obvious heuristic algorithm. (We mention in passing that this claim holds for *all* sets $A \subsetneq \Sigma^*$.) This of course does not say that *every* NP-complete problem has, for *every* $k$, some polynomial-time heuristic algorithm whose error frequency is $2^{n^{1/k}}$; the previous section discusses how Buhrman and Hitchcock shown that that would induce highly unlikely consequences.[2]

Also, it is important to keep in mind that although in this article we have been focusing completely on the simple case of just counting the number of errors, in some sense the above padding trick makes the point that this approach has some shortcomings: the frequency can be changed sharply by simple actions, such as many-one reductions, that are generally in other contexts considered rather innocent, complexity-theoretically. (The desire to be protected from being changed sharply by actions that one would hope would not result in changes is related, in a slightly different context, to why Levin's notion of average polynomial time is defined in a way that is more subtle and transformed than just directly averaging running times [Lev86].) Nonetheless, since many people in the real world find error frequency a natural thing to discuss, it is reasonable and important to look at error frequency.

This is also a good time to mention that various notions of heuristic polynomial time have long been studied that themselves are defined in terms of flexible probability distributions over the inputs, and also that it long has been well-known that it is important not to confuse the notions having heuristic polynomial-time algorithms with the notion of having the average of one's time complexity be good. See, for example, [Tre02, PR07, Imp95, BT06, EHRS09]. Readers interested in a more complete view of typical-case and average-case hardness issues (and we mention in passing that the latter term is often used to refer to the former rather than the different notion that Levin intended by the latter) should certainly look at those sources, for a broader and deeper treatment than is given here where our focus has been on presenting what was achieved by the early 1990s, and most particularly we mention the book by Bogdanov and Trevisan [BT06]. Also particularly interesting are the various worlds put forward in the early, playful, influential paper of Impagliazzo ([Imp95], see also Impagliazzo's recent relativized advance [Imp11] related to that earlier program).

## 2.3 The 1970s, Berman-Hartmanis, and Inheritance of Frequency of Correctness: If I'm Frequently Easy, Then So Are All These Famous Problems

Section 2.1 sections mention results yielding complexity collapses if NP-hard sets have error frequencies that are too low. Those results were proven based on a very well-defined line of complexity-theoretic research, namely, the study of whether there can be sets that are hard (with respect to various types of reductions) for NP yet that are sparse (or at least are of reasonably low density).

---

[2]Furthermore, if an NP-complete problem like SAT had a polynomial-time heuristic with error frequency bounded by $2^{n^\varepsilon}$ for all $\varepsilon > 0$, then we could conclude that all NP problems have subexponential-size circuits, which also is thought unlikely.

However, though that line of research took on a life of its own, in historical context it was started to address an even earlier issue, namely, the Isomorphism Conjecture. In particular, Berman and Hartmanis [BH77] in the 1970's boldly conjectured that all NP-complete sets were (polynomial-time) isomorphic—that there really was only one NP-complete set, and it simply appeared in a variety of disguises. To give this teeth, they proved that essentially all then-known NP-complete sets were isomorphic to each other. However, SAT is dense, and dense sets cannot be isomorphic to sparse sets. So one potential line of attack on the Isomorphism Conjecture would be to construct an NP-complete sparse set. This led to intensive research into that issue, and eventually Hartmanis's student Mahaney famously and beautifully proved that if there is a sparse NP-hard set then P = NP ([Mah82], see also the surveys [Mah86, You92, HOW92, GH00] for a broader view of what came before and after in the study of sparse hard sets). Thanks to Mahaney's work, the sparseness-based line of attack on the Isomorphism Conjecture was revealed as being unpromising: To make it work, one would need to collapse NP to P. However, it is clear that if NP equals P then the Isomorphism Conjecture fails, since in that case there are finite NP-complete sets. So Mahaney showed that the sparseness line of attack is no easier than the already then-known (and far from easy, one would think!) attack line of shattering the Isomorphism Conjecture by proving that P = NP.

The reason we mention all this is that it has recently been noticed that the tools that Berman and Hartmanis [BH77] built to support their result that all then-known NP-complete sets are mutually isomorphic also give consequences in terms of how often sets can be hard. Results in this direction can be obtained not just from the "sparse hard sets" line of work that grew out of an issue raised by Berman and Hartmanis, but can also be obtained by using the techniques *of* Berman and Hartmanis—though the flavor of the results that one gets is somewhat different.

The observation we are speaking of is made in a recent paper of Hemaspaandra, Hemaspaandra, and Menton [HHM12]. We won't go into great detail here, but rather will just mention the flavor of their observation. They make the observation because they need to argue about the error frequency of some sets related to NP ∩ coNP, a class for which the standard tools and results that exist for NP aren't available. However, as they note (in their footnote 13), the same observation that they make will also work for NP. What they observe is that if there exists even one NP set that is frequently hard (i.e., on which no polynomial-time heuristic algorithm has a low error frequency), then every single set that is NP-hard with respect to polynomial-time *one-to-one* reductions will be essentially (at least) just as frequently hard, give or take a slight bit of flexibility (more particularly, the argument of the error frequency lower bound changes from $n$ to $n^\varepsilon$, where here we are speaking of error-frequency counts among the strings of length up to and including $n$).

Now, the reason why that is interesting is that a central part of the work of Berman and Hartmanis [BH77] was—using a close study of paddability—that they showed that essen-tially all famous NP-complete problems were not just polynomial-time many-one complete for NP, but were in fact polynomial-time *one-to-one* complete for NP.

Putting that all together, what one has is that if any NP set, even some crazy, artificial one, is frequently hard, then all the famous NP-complete sets are (at least) almost as frequently hard as that crazy set. In the contrapositive, this says that if any of the famous, natural NP-complete sets is not very frequently hard, then all of NP is not very frequently hard. When [HHM12] discusses how this can be squared with the general belief that some natural NP-complete problems have heuristic algorithms that seem to do a spectacular job, that paper points to the same density discussion that we've included above: The slightly flexibility in the hardness lower bound is enough to shift densities down from nearly all strings to just around $2^{n^{1/k}}$ strings, and so if one believes that each natural NP-complete set for some $k$ will (for each polynomial-time heuristic algorithm) on infinitely many lengths have about that many hard strings, one isn't trapped in an intellectually inconsistent position. However, if for example one believes there are polynomial-time heuristic algorithms that err on SAT on at most quasi-polynomially many strings, then one must also believe in polynomial-time heuristic algorithms for each NP set that err on at most quasi-polynomially many strings.

To summarize Section 2: Work from the early 1970s through the early 1990s [BH77, Sch86, OW91, BH92, HL94] explicitly or implicitly showed that if NP-hard sets have (deterministic) polynomial-time heuristic algorithms that are very infrequently incorrect, then strong, unlikely complexity-theoretic class collapses and consequences follow.

## 3 Deterministic Heuristic Algorithms for Problems Solvable With Randomness

So far, we have seen that when an efficient algorithm solves an NP-hard problem on all but a small number of inputs (i.e., it has very "low error"), then there are also efficient algorithms with "no error," i.e., that solve the same problem on *every* input. This connection is useful for ruling out heuristics that seem too good to be true. What about problems where we believe there *is* an efficient algorithm? Perhaps we can make progress on designing algorithms for these problems by first designing heuristics, and then (finding and) applying some theorem that, analogously to what Corollary 2.2 did for NP-hardness case, allows us to indirectly remove the remaining errors. In this section, we will discuss a situation like this that is central to modern complexity theory: simulating randomized computations with efficient deterministic algorithms.

A long and significant line of work has led the majority of complexity theorists to conjecture that randomized polynomial time (by which we here mean bounded-error probabilistic polynomial time [Gil77]) is really no different from deterministic polynomial time; that is, it is generally believed that P = BPP.[3] But we seem to be far from proving

---

[3]Why do complexity theorists believe this? The short answer is that strong circuit lower bounds actually imply P = BPP, and complexity theorists currently are more inclined to believe the circuit lower bounds

P = BPP, for various reasons (one of which is that P = BPP itself would imply new circuit lower bounds [KI04]). Can we design deterministic heuristics for BPP with low-error frequency? This topic was discussed more generally in an earlier *SIGACT News* article by Shaltiel [Sha10]. Here, we will briefly survey the subset of this work that helps provide the best contrast between the issues of deterministic heuristics for NP described earlier, and deterministic heuristics for BPP.

In a paper entitled "Derandomization that is rarely wrong from short advice that is typically good," Goldreich and Wigderson [GW02] use an existing pseudorandom generator to show that, for every $\varepsilon > 0$, there is a LOGSPACE algorithm for connectivity in undirected graphs that is correct on all but $2^{n^\varepsilon}$ of the $n$-node graphs. (Note this is an unconditional result—it does not rely on any hardness assumptions.) They also prove a similar result for simulating BPP with a heuristic algorithm (under some plausible circuit lower bound assumptions): for every $\varepsilon > 0$, every BPP algorithm can be simulated by some deterministic polynomial-time heuristic that errs on at most $2^{n^\varepsilon}$ of the $n$-bit inputs.[4] Their key insight is that, when a randomized algorithm requires only a small amount of randomness, *the input itself can be seen as a source of randomness*. By using the input to seed a pseudorandom generator they can develop, for any desired $\varepsilon > 0$, a heuristic algorithm that works on all but at most $2^{n^\varepsilon}$ of the $n$-bit inputs.

Judging from our earlier discussions on turning low-error heuristics into no-error algorithms for NP-hard problems, one might predict that there should be LOGSPACE algorithms for connectivity on *all* graphs. It turns out that this prediction is accurate! Reingold [Rei08] gave an amazing LOGSPACE algorithm for s-t connectivity that works on all graphs, and the algorithm can be seen as a certain kind of pseudorandom generator in itself.

Several other papers build on the ideas of Goldreich and Wigderson. Van Melkebeek and Santhanam [MS05] weaken the "plausible circuit lower bound assumptions" used to obtain a heuristic simulation of BPP in P: informally, their assumption is that for all $k$, there is a P set that cannot be heuristically simulated by Merlin-Arthur machines running in time $n^k$. (The negation of this assumption says something like "every P set can be simulated by Merlin-Arthur machines in $n^{100}$." This sort of arbitrary speed-up of P seems unlikely.) Zimand [Zim08] gave deterministic heuristic algorithms for simulating sublinear-time randomized algorithms. For example, if a set $S$ can be decided in $n^{o(1)}$ time with randomness, then there is a heuristic algorithm running in $n^{o(1)}$ time without randomness (assuming a random-access model of computation) that is wrong on at most $2^{n-n^{o(1)}}$ inputs

---

than we are to believe that randomness is powerful. A slightly longer answer is that there are many papers showing that various plausible lower bounds imply nontrivial simulations of randomized computation, so there does seem to be a collection of evidence in favor of BPP being deterministically easy.

[4]The reader may wish to call a foul here. After all, this is setting the error rate to be at most $2^{n^\varepsilon}$. And in the previous section, we mentioned that by padding it is simple to make even NP-complete problems that are easy on all but $2^{n^\varepsilon}$ of their inputs. However, there is a key difference that distinguishes the cases. For the NP case, different $\varepsilon$ values require different sets; as $\varepsilon$ drops we have to pad more energetically. For the BPP case, the claim is the much stronger one that for each BPP set and for every $\varepsilon > 0$, we can build a heuristic algorithm that errs on at most $2^{n^\varepsilon}$ length $n$ inputs. So there is no foul here; there is a difference in how strong and sweeping the claims are.

of length $n$. In this case, the input is much longer than the running time and that makes for a rich source of randomness.

Perhaps there is a more general connection behind these results. Generally speaking, when does "derandomization that is rarely wrong" entail "full derandomization"? We know that, for NP-hard problems, heuristics that are wrong on few inputs entail completely correct algorithms; a similar generic connection for BPP problems would be extremely interesting, and useful, given the "rarely wrong" simulations that we already know. We do know that one significant consequence of fully-correct derandomization can still be accomplished with only heuristic algorithms. In a celebrated paper, Kabanets and Impagliazzo [KI04] showed that deterministic subexponential-time algorithms for polynomial identity testing (more precisely, testing whether an arithmetic circuit is identically zero over $\mathbb{Z}$) implies circuit lower bounds for nondeterministic exponential time. Kinne, Van Melkebeek, and Shaltiel [KMS12] have shown that the deterministic algorithm can be replaced with a heuristic in that theorem and the result remains true: a $2^{n^{o(1)}}$-time algorithm (for arithmetic circuit identity testing) that is correct on all but $2^{n^{o(1)}}$ arithmetic circuits of size $n$ still implies circuit lower bounds for NEXP. This suggests that the difficulties faced in trying to prove theorems like P = BPP still persist when trying to prove BPP merely has deterministic polynomial-time heuristics with few errors. So the two objectives may well be equivalent.

In this section, as in the others, we have neglected to mention a great deal of important work in derandomization, simply because the error bounds are "too large" for the focus of this article. This work is covered in Shaltiel's survey [Sha10].

## 4 When Heuristics Fail Often in Theory But Work Well In Reality

In this last part of our article, we will take a different look at heuristics for hard problems. Suppose we resign ourselves to the idea that we won't be able to solve almost all instances of NP-hard problems in polynomial time (as Section 2 strongly suggests). This resignation must somehow be reconciled with an empirical fact: many, many instances of NP-hard problems generated in practice *can* be solved, often even by heuristics that are provably *bad* on almost all instances!

The SAT problem on conjunctive normal form (CNF) formulas is such a problem. Modern SAT solvers need exponential time to solve random kCNF instances.[5] So, if they were restricted to run in polynomial time (e.g., forcing a *Don't Know* answer if no definite answer has been found after polynomial time), these solvers would have a very high error frequency. However, they still manage to solve a huge variety of instances that come up in industrial settings. (One recent example: SAT solvers were used to help validate portions of the Intel Core i7 processor [KGN$^+$09].) The practical formulas from industry do not behave at all like "average" instances from the space of *all* instances; evidently, they come from a very restricted and highly structured subset of the SAT problem. If we could accurately model what this subset looks like, we could focus more of our energy on solving it, and better understand how to improve the performance of SAT solvers in practice.

---

[5]Here, we consider the usual distribution of kCNF instances drawn at random, where the ratio of clauses to variables (given by the parameter $\Delta$) is a fixed constant. For instance, the propositional proof system that arises from algorithms such as DPLL (Davis-Putnam-Logemann-Loveland) with clause learning can be efficiently simulated in the good old Resolution proof system (this is basically folklore), and it is known that Resolution requires proofs of $2^{\Omega(n)}$ size on random CNF formulas (see [BW01, BKPS02] or the more recent survey of Segerlind [Seg07]). Furthermore, local search algorithms are only known to take super-polynomial time on random kCNF formulas in general (but for small enough clause densities there are polynomial-time local search algorithms—see [AB07]). Special-purpose algorithms designed to work well on random instances can fare better, but still are not provably efficient for all ratios $\Delta$. (The conjecture that *no* polynomial-time algorithm exists that correctly determines the satisfiability of almost all random 3CNF formulas, regardless

Here we propose theoretical models that attempt to capture this real-world phenomenon: one based on the "compressibility" of real-world SAT instances and one based on the "closeness" of real-world SAT instances to efficient special cases of SAT. The discussion will be extremely open-ended, and there are many natural questions related to these models. We much hope that the readers of this *SIGACT News* article will answer some of them (or perhaps propose new or better questions).

## 4.1 Compressible Instances

One key property of the formulas from practice is so obvious that it can easily be overlooked: these instances are the *outputs of some reduction*. The inputs to this reduction are problem instances from some application domain, with many different constraints and logical restrictions on what constitutes a solution. There will also likely be many *redundancies* within the problem encoding. For example, if the instance encodes a potential design for a digital circuit, then informally speaking, various pieces of that circuit will have been modularly designed and will exhibit many redundant subparts. The outputs of the reduction are simply CNFs that faithfully encode all these constraints, and because all the underlying logic in the real-world instance must be modeled in the CNF, these outputs are often considerably larger in size than the respective input, and the CNF encoding itself often introduces even more redundancies and symmetries into the problem. (For instance, the so-called "frame problem" can arise: suppose some entity may be in location $i$ at time $t$, we have a variable $x_{i,t}$ for that assertion, and the entity does not "move" during step $t+1$. Then we need to assert that $(x_{i,t+1} \iff x_{i,t})$.)

All of this suggests that, due to the way in which complex systems are designed in practice (hardware, software, space shuttles, etc.), the CNF formulas that encode the problems of verifying these designs will be highly *compressible*, meaning that there is an efficient process with a relatively short representation that can produce the clauses of the CNF formula. Note that logistics/planning/scheduling problems also naturally have underlying redundant structure, as the various entities being planned/scheduled are often quantified over a small number of types. Just to give an example of what we mean: in airline scheduling, the number of different *kinds* of airplanes is small compared to the number of different *routes* that must be scheduled. It follows that, in a formula that models an airline's complete flight

---

of the constant $\Delta$, is essentially the Random 3SAT Hypothesis of Feige [Fei02, Bar12].)

schedule, the kind of logic used for describing airplane types will be replicated many times, over the many routes that must be taken.

Given the above points, it would appear that the computational problem being solved in the real world is not the generic all-purpose SAT problem familiar to theoreticians, but rather a "compressible" SAT problem, where there exists a short representation of the given CNF formula, and we wish to know if the CNF is satisfiable. This scenario can be naturally formalized in a complexity-theoretic way with the following problem.

> SUCCINCT SAT: Let $x_1, x_2, \ldots, x_{2^n}$ be the $n$-bit strings in lexicographical order. Given a Boolean circuit $C$ with $n$ inputs, does its truth table $T = C(x_1)C(x_2)\cdots C(x_{2^n})$ encode a satisfiable CNF formula?[6]

So, we're modeling the general compression of a redundant formula using Boolean circuits; the circuit $C$ provides a short representation of the formula. The choice of using truth-tables of circuits in our representation captures the fact that the various pieces of real-world formulas are typically largely independent of other pieces. When we want the $i$th bit of the CNF formula, we do not have to "decompress" the entire formula to get it: we can simply evaluate $C$ at the input $x_i$. (Moreover, Boolean circuits are a natural, flexible, and compact way of holding information and expressing finite functions.)

What can be said about the complexity of SUCCINCT SAT? Surely the problem got easier, as we're only focusing on compressible instances of SAT. Actually, the problem got harder, as the following shows.

**Theorem 4.1 (Papadimitriou [Pap94], pp. 494–495)** SUCCINCT SAT *is* NEXP-*complete under many-one polynomial-time reductions.*

What's going on? Well, our input changed: rather than the formula, we're given a circuit representing a compressed version of the formula. It could happen (and it does) that the compressed version of the formula is exponentially smaller than the formula, yet the resulting formula still captures a difficult, expensive computation. This is the intuitive reason why SUCCINCT SAT is NEXP-complete.

The situation is not so bad: if our formula is really exponentially larger than our representation for it, then it's OK to take exponential time to solve it. (That would only translate to polynomial time, or perhaps quasi-polynomial time, in the length of the formula.) So measuring the time complexity by the size of the circuit encoding the formula is somewhat misleading; what we want to measure is the time complexity for solving the final formula obtained. If we can design an algorithm for such formulas that runs in significantly less than $2^n$ time, where $n$ is the number of variables to the formula, then we'll have explained (in some small way) why formulas that come from practice get solved so efficiently.

---

[6]Here, we are totally neglecting what the encoding of the CNF formula looks like; for the purposes of our discussion, any reasonable encoding of CNF (where, for example, all bits pertaining to a certain clause appear consecutively in the encoding) will suffice.

However, solving even highly compressible instances efficiently can still lead to resolving major open problems in complexity theory. Let us say that an $n$-bit formula $F$ is *super-compressible* if there is a $\log^{\mathcal{O}(1)} n$-size circuit whose (induced) truth-table is equal to $F$. If the satisfiability of all super-compressible formulas can be determined in $n^{\log^{\mathcal{O}(1)} n}$ time, then we'll have proved NEXP = EXP. This is because an $N^{\log^{\mathcal{O}(1)} N}$ time SAT algorithm, in terms of the number of $N$ variables of the formula, is still $2^{n^{\mathcal{O}(1)}}$ time in terms of the circuit $C$ of size $n$, and our Succinct SAT problem is NEXP-complete. Therefore a "quasi-polynomial" time algorithm for SAT on compressible formulas with $N$ variables implies NEXP = EXP.

In fact, our circuit representation could be extremely restricted, even with bounded depth (i.e., $AC^0$), and yet the SAT problem for formulas encoded by small $AC^0$ circuits is still NEXP-complete. This was essentially observed by Arora, Steurer, and Wigderson [ASW09], who studied succinct representations for NP-complete graph problems, showing that (for example) the following problem is NEXP-complete.

> Succinct MAX-CLIQUE: Given an $AC^0$ circuit $C$ and integer $K$, does the truth-table of $C$ encode the adjacency matrix of a graph with a clique of size $K$?

It seems plausible that MAX-CLIQUE can be solved in $2^{o(n)}$ time on $n$-node graphs that have $\log^{\mathcal{O}(1)} n$-size $AC^0$ circuits encoding them, simply because we know that $AC^0$ circuits are so restricted (they can't compute PARITY, for example [FSS84]). This would "only" imply inclusions like NEXP $\subseteq$ TIME$[2^{o(2^n)}]$, which could well be true, and very interesting.

Suppose for the moment that we posit the following (somewhat imprecise) thesis.

> **Succinct SAT Thesis:** "SAT in practice" corresponds to Succinct SAT, where each size-$\ell$ circuit (within the framing of Succinct SAT) is now additionally required to have $\log^{\Omega(1)} \ell$ inputs.

That is, "SAT in practice" is a subset of Succinct SAT, where size-$\ell$ circuits "decompress" to formulas of size $2^{\log^{\Omega(1)} \ell}$. So we require that each formula we care about exhibits some considerable level of compression: from an $\ell \leq 2^{\log^{\delta} m}$ size circuit to an $m$-length formula. According to this thesis, if we believe that EXP $\neq$ NEXP (a stronger belief than P $\neq$ NP, but still believed) then we still shouldn't believe that *all* of "SAT in practice" can be solved in polynomial time. On the other hand, perhaps "SAT in practice" can be solved in time *subexponential* in the number of variables, as suggested above with Succinct MAX-CLIQUE. Then we would have inclusions resembling NEXP $\subseteq$ TIME$[2^{o(2^n)}]$.

The true test of the Succinct SAT thesis is whether it can be used to predict the performance of practical SAT solvers. One could try to strengthen the thesis.

> **Strong Succinct SAT Thesis:** "SAT in practice" is easy because these instances are Succinct SAT instances.

This thesis would suggest that the Resolution proof system (which can model essentially all variants of the DPLL algorithm that we know to be used in practice) should yield relatively short proofs of unsatisfiability for many compressible formulas that are unsatisfiable. Note that there *are* some "simple" unsatisfiable formulas (such as "Pigeonhole Principle" formulas) that require exponentially large Resolution proofs [Hak85, Raz04], and have very small circuit representations.[7]

There is a parallel in the real world: modern SAT solvers can also struggle to solve Pigeonhole formulas, unless they introduce "symmetry-breaking" techniques to solve them (see [GKSS08], Section 2.2.5). In fact, currently the "best" modern SAT solvers are portfolios of many other solvers, which have been trained using machine-learning techniques to recognize when various structured things like Pigeonhole formulas are being handed to them (and run the appropriate solver on those formulas).[8] So these theses seem safe for now; what would really challenge them are formulas that arise naturally in practice, are easy for modern solvers, but are highly incompressible. The only formulas we can presently think of that are highly incompressible and in high demand to be practically solved are those that stem from integer factorizations (of numbers that are the products of two random large primes). However, these instances are not among those that are presently easy for solvers (otherwise, we would be very worried about cryptographic security).

We conclude this section by suggesting that readers try to falsify or give more evidence for these hypotheses about SAT. Even if they are close to the truth, how can we use these theses to better explain the performance of modern SAT solvers? One potential intuition for answering this question is that sub-double-exponential-time algorithms for NEXP problems are no less likely, and perhaps more likely, to exist than subexponential-time algorithms for NP problems, so we may expect theoretically that compressible instances are possibly easier to solve than incompressible ones. Other intuitions will be mentioned in the next section.

## 4.2 The Distance from Practical Instances to Theoretically Easy Instances

The second model we wish to discuss was proposed by Williams, Gomes, and Selman [WGS03], and it has a subtle relationship with the Succinct SAT ideas. Those authors' motivating thesis was that modern SAT solvers do well because of two factors:

(a) the instances in practice are "close" to subsets of SAT that are worst-case "easy," and

(b) the solvers in practice implicitly and efficiently reduce given instances to one of these "easy" subsets, which is possible because of (a).

---

[7] Recall the Pigeonhole formulas (falsely) assert that $n+1$ pigeons can fit in $n$ pigeon-sized holes. For $i = 1, \ldots, n+1$ and $j = 1, \ldots, n$, there is a variable $x_{i,j}$ that is true if and only if pigeon $i$ is in hole $j$. To express that the $i$th pigeon maps to a hole, we add the clause $(x_{i,1} \vee \cdots \vee x_{i,n})$. To express that pigeons $i$ and $i'$ cannot both fit in hole $j$, we add the clause $(\neg x_{i,j} \vee \neg x_{i',j})$. Given the extreme redundancy of this formula, one can derive a small circuit of $\log^{\mathcal{O}(1)} n$ size that can print arbitrary bits of these clauses on demand. These circuits are therefore "no"-instances of SUCCINCT SAT.

[8] One example of such a solver is SATzilla [XHHL08]. See http://www.satcompetition.org/ for more information on the SAT solver competitions.

This informally explains why randomly restarting a SAT solver can often yield large gains in performance: by restarting a solver often, one is forcing it to spend more of its time trying to perform a quick reduction to an "easy" instance (computing this "distance" to an easy instance), rather than spending time exploring deeply in the search space.

More formally, the authors gave a formal concept of a *backdoor set of variables* to a formula. Let $h$ be a polynomial-time errorless heuristic (which answers *SAT*, *UNSAT*, or *Don't Know*, on each instance). A backdoor set of variables for a formula $F$ is defined with respect to $h$ to be a subset $S$ of the variables of $F$ such that once the variables of $S$ are all assigned to some values, the remaining formula $F'$ is decided by $h$ (that is, $h$ returns a *SAT* or *UNSAT* answer).[9] A small backdoor set for $F$ indicates that $F$ is hiding considerable structure inside: by setting just a few variables in $F$, the resulting formula becomes polynomial-time solvable! Hence, $F$'s "distance to tractability" is surprisingly small.

For every DPLL-based solver, one can extract a polynomial-time errorless heuristic (it is essentially what one obtains by turning off all branching/backtracking in the solver, and forcing the solver to output *Don't Know* if it decides to branch on a variable). It has been empirically demonstrated over the years that many real-world instances have very small backdoors with respect to these heuristics, and the variable-choice heuristics used by solvers are good at selecting variables within these backdoors (some references are [WGS03, DGS07, SS08, DGS09, DGM+09]). The theoretical study of backdoor sets has also received significant attention (see the survey of Gaspers and Szeider [GS12]).

We stress that the existence of small backdoor sets is *not* tautological: we have not defined them in such a way that they *must* exist when a SAT solver runs efficiently on some formula. Their existence implies that satisfying assignments can be found, in principle, at shallow levels of the "recursion tree" representing a DPLL solver's branching. But this is not true in general for combinatorial search problems solved by similar methods. One dominant class of methods for solving integer programs is *branch-and-bound*: one relaxes the integer program to a linear program over the rationals, and recursively adds new inequalities to the program (say, $x \leq 1$ in one branch, and $x \geq 2$ in the other), so that (a) the program remains feasible and (b) the remaining solutions will be eventually forced to be integral. Conventional wisdom in operations research says that most integral solutions are found only *deep* within this recursion tree, not at shallow levels (for a reference, see [Van01, p. 391]). So the observation that satisfying assignments in practice are often found at shallow levels is important. It calls into question whether *depth-first* search techniques like DPLL-based methods are really the best ways to attack many practical instances. Indeed, as mentioned above, small backdoors help explain why DPLL solvers augmented with rapid random restarts can be so effective [GSK98].

From the above observations, one can readily pose theses concerning the existence of small backdoor sets in practical instances. How do such theses square with the Succinct SAT Thesis? There is certainly a connection here. The Succinct SAT Thesis implies that practical formulas are compressible in such a way that they can be efficiently decompressed. A small backdoor set for a formula $F$ yields a way of "compressing" a *satisfying assignment* to $F$: given the small set of variables, some particular values for them, and $F$ itself, one can efficiently recover some satisfying assignment to $F$. Furthermore, when a formula $F$ does possess a small backdoor set, this indicates $F$ itself is quite *nonrandom*. Interian [Int03] showed that, with respect to certain natural heuristic algorithms, a randomly drawn 3-CNF formula will (with high probability) have only backdoor sets of size $\Omega(n)$. Perhaps the Succinct SAT Thesis predicts the existence of small backdoors in practice? That is, perhaps compressible SAT formulas should have compressible SAT assignments.[10] For example, take the set of CNF formulas generated by tiny $AC^0$ circuits; must they necessarily have a small backdoor with respect to some natural heuristic algorithm? We do not yet know how to answer questions like that, but suspect that some form of such questions can be answered affirmatively. What seems to be true is that many practical instances are compressible in an efficient way, and when they are satisfiable, they possess at least one satisfying assignment that is also compressible in a way that makes that assignment relatively easy to find.

---

[9] One can give an even more formal definition of what particular characteristics the heuristic algorithm should have, but for the purposes of this discussion we omit it.

The backdoor concept gives some insight into why solvers do well on real-world formulas, but its explanatory power goes only so far. One begs to ask: *why are the backdoor sets there in the first place?!* Yes, there is "structure" in these instances, but why is it manifested in this way, in the real world? Is it because these SAT instances are compressible outputs of other reductions, or is there an even deeper explanation? In principle, a serious scientific explanation of the practical SAT-solving phenomenon could require positing axioms that are rooted in social science. It could be that small backdoor sets arise because of the structured ways in which people design complex systems in practice, and the design patterns they apply lead naturally to compressible instances with small backdoor sets.

One final thought: small backdoors in instances verifying the correctness of software and hardware are of course a positive aspect for *verifying* these systems, but their presence can also indicate *inefficiencies* in the designs of these systems. (Indeed, the speed of modern SAT solvers can also be used to quickly find security exploits as well! See [CGP+08] for an example.) Theory would predict that we must be missing a wide range of possibly more efficient and more secure software designs, if all the software we verify in practice has such extreme structure.

**Acknowledgments**  We are grateful to Paul Beame for providing helpful SAT references.

# References

[AB07]  M. Alekhnovich and E. Ben-Sasson. Linear upper bounds for random walk on small density random 3-CNFs. *SIAM Journal on Computing*, 36(5):1248–1263, 2007.

---

[10]This is probably not true in its strongest possible form, as that would imply NEXP = EXP. (Given a SUCCINCT SAT instance, we could try all small circuits to see if any encode a satisfying assignment for the instance.)